\documentclass[]{PHMSociety}

\usepackage{graphicx}
\usepackage{amsmath,amssymb,amsfonts}
\usepackage{xcolor}
\usepackage{booktabs}
\usepackage{apacite}

\begin{document}

\title{On Adversarial Vulnerability of PHM algorithms – An Initial Study}

\author{%
	Weizhong Yan\authorNumber{1} , Zhaoyuan Yang\authorNumber{2}, and Jianwei Qiu\authorNumber{2}
}

\address{
	\affiliation{{1}}{AI \& Machine Learning, GE Research Center, Niskayuna, New York, 12309, USA}{ 
		{\email{yan@ge.com}}
		} 
	\tabularnewline 
	\affiliation{2}{Computer Vision, GE Research Center, Niskayuna, New York, 12309, USA}{ 
		{\email{zhaoyuan.yang@ge.com}}\\
		{\email{jianwei.qiu@ge.com}}
		} 
}

\maketitle

\phmLicenseFootnote{Weizhong Yan}

\begin{abstract}
With proliferation of deep learning (DL) applications in diverse domains, vulnerability of DL models to adversarial attacks has become an increasingly interesting research topic in the domains of Computer Vision (CV) and Natural Language Processing (NLP). DL has also been widely adopted to diverse PHM applications, where data are primarily time-series sensor measurements. While those advanced DL algorithms/models have resulted in an improved PHM algorithms' performance, the vulnerability of those PHM algorithms to adversarial attacks has not drawn much attention in the PHM community. In this paper we attempt to explore the vulnerability of PHM algorithms. More specifically, we investigate the strategies of attacking PHM algorithms by considering several unique characteristics associated with time-series sensor measurements data. We use two real-world PHM applications as examples to validate our attack strategies and to demonstrate that PHM algorithms indeed are vulnerable to adversarial attacks.

\end{abstract}

\section{Introduction}
Prognostics and health management (PHM) is a modern maintenance strategy that aims for reducing operation and maintenance (O\&M) costs by reducing unscheduled repairs and increasing availability of industrial assets.  PHM involves several technical components or predictive algorithms/models, including fault detection, fault diagnosis, fault prognosis, and logistical decision-making based on predictions \cite{phm_definition}. The predictive accuracy and robustness of those predictive models are the keys to enabling PHM to achieve maximal business values. 

Towards achieving the highest predictive accuracy of the PHM models, deep learning (DL), regarded as a state-of-the-art ML technique, has been increasingly adopted in PHM applications in recent years, for example, \cite{LIN_2018555, yan_detection_2019}. \cite{FINK_2020} and \cite{Zhang_2019} provided a broad review of different deep learning techniques used in diverse PHM applications.

As shown in the domain of computer vision, deep learning models are vulnerable to adversarial attacks\cite{IPNN,fgsm,moosavi2017universal,su2019one,eykholt2018robust}. That is, small deliberately-designed perturbations to the original samples can cause the DL model to make false predictions with high confidence scores. DL models' vulnerability to adversarial attacks is well studied in CV. However, to date adversarial attacks on PHM algorithms or, more generally, PHM solution security have not been actively studied.

For a majority of PHM applications, the data used by PHM models are predominantly multivariate time-series sensor measurements, as opposed to 2D images in computer vision. The time-series sensor measurement data has it own unique characteristics, including:  1) noisy and unreliable due to the faulty or failed sensors, 2) multimodal and heterogeneous, i.e., data coming from different types of sensors, e.g., temperature, pressure, accelerometers, and 3) having strong  spatio-temporal dependencies. These unique characteristics associated with time-series sensor data pose additional challenges and require special design strategies in attacking as well as defending PHM algorithms. 

On the other hand, the economic impact of adversarial attacks to these PHM solutions can be significant or even bigger than that to hard perceptual problems, simply because most of PHM applications involve safety-critical and time/cost-sensitive industrial assets, e.g., power grids, power plants and gas turbines. Take fault detection as an example. Fault detection is to detect problems and abnormal behaviors of assets or processes earlier to prevent catastrophic damages. An adversary in this case can manipulate the time-series data to cause the detection algorithms to miss-detect the faults or failures on time, thus resulting in catastrophic damages to the machine. Alternatively, an adversary can simply temper the normal time-series data to force the algorithm to generate a large number of false alarms, which results in unnecessary manpower for tracing the false alarms and even unnecessary machine downtime. 

Despite the aforementioned importance, securing PHM solutions from adversarial attacks has been largely ignored yet. Very recently, Zhou et al. \cite{zhou2020overcoming} demonstrated that deep learning prognostics models are vulnerable to adversarial attacks. To the best of our knowledge, work by Zhou et al. was the only one related to PHM algorithms' security.  With proliferation of PHM solutions deployed for a large variety of mission-critical industrial assets, more active research efforts are in great need on developing proper strategies for attacking as well as defending PHM solutions by considering the specific characteristics of time-series sensor measurements data involved in PHM applications.

Motivated by these needs, in this paper, we study vulnerability of PHM algorithms where time-series sensor measurements are the primary data type. More specifically, we explore attack strategies by exploiting the unique characteristics of time series sensor data. We use two real-world PHM applications to validate the attack strategies and to demonstrate that PHM algorithms are vulnerable to adversarial attacks.   

The rest of the paper is organized as follows. Section 2 reviews related work. Strategies and attack model details are discussed in Section 3. Section 4 presents experiments and their results, while Section 5 concludes the paper.

\section{Related Work}

In the past a few years, adversarial machine learning (AML) has emerged as a hot research topic (\cite{10.1145/3134599,Kianpour2019TimingAO}). There are various adversarial attack methods for deep learning models. Inference-time attack and the training-time attack are two common adversarial attacks for neural networks. For the inference time-attack, an adversary adds small perturbations to input measurements so that a machine learning model produces incorrect predictions with high confidence~\cite{fgsm, IPNN, carlini2017towards, kurakin2016adversarial}. Later,~\cite{moosavi2017universal} demonstrate a way of generating an universal adversarial perturbation for a trained classifier,~\cite{su2019one} show an approach of generating one-pixel attack against a classifier. Most of attacks are generated in the digital domains by manipulating the digits of an image,~\cite{eykholt2018robust} demonstrate that this type of attack are also feasible in the physical world.
For the training-time attack, training data are corrupted with carefully designed backdoors or triggers~\cite{purdueTrojan}. Through injecting the backdoor into the training data, the poisoned models will make false predictions~\cite{badnets}. In this paper, we focus on inference-time attack and will demonstrate the attack using experimental data from two PHM applications.

Most of the adversarial attacks are demonstrated in CV and NLP applications~\cite{IPNN, fgsm, papernot2016crafting}. For example,~\cite{fgsm} uses the fast gradient sign method (FGSM), and~\cite{papernot2016crafting} uses the forward derivative method to craft adversarial examples. More recently, \cite{fawaz2019adversarial} use the FGSM methods on time series classifications to investigate the adversarial attacks on the vehicle sensor and food data classification problems. Adversarial attacks on PHM solutions/algorithms have not been actively studied. Very recently, \cite{zhou2020overcoming} demonstrated adversarial attacks on Remaining Useful Life (RUL) of turbo fan engines. To the best of our knowledge, this is the only work that related to our work in this paper.

\section{Adversarial attacks on PHM algorithms} \label{S:attacks}

In this section, we describe our strategies of attacking PHM algorithms. PHM solutions generally have four categories of functional algorithms, namely, fault detection, fault diagnosis, fault prognostics and logistic decison-making. While detection and diagnosis are a classification problem, prognostics is a regression problem and logistic decision-making is an optimization problem. In this paper, we focus on strategies on attacking fault detection and prognostics algorithms and leave adversarial attacks of logistic decision-making to our future work.

\textbf{PHM algorithms attack scenarios considered}  PHM models perform their functionalities based on the sensor measurements from the asset monitored. These sensor measurements typically are communicated to PHM models via a communication protocol. We assume the attacker can access the communication channels and thus can attack the PHM solutions by manipulating the sensor measurement signals. We also assume the attacker has the full knowledge of the PHM algorithms, that is, in this paper we consider white-box attacks (\cite{Yuan19TNNLS}).

An attacker can attack the PHM solutions at either training time (training-time attacks, also called ``poisoning attacks") or inference time (inference-time attacks, also called ``evasion attacks"). In this paper we only consider inference-time attacks. The inference-time attack refers to an adversarial attack in the inference stage after a model is built and deployed. Inference attacks can be targeted attack and non-targeted attack~\cite{Yuan19TNNLS}. We do not limit our attack to a particular class, thus we generate adversarial examples with the more general non-targeted attack.

\textbf{Problem formulation} Let $\mathbf{x}=(x_1, x_2,...,x_n) \in \mathbb{R}^{n}$ be the multivariate time-series sensor measurements, the input signals to PHM algorithms. We can formulate an adversarial attack as an optimization problem as shown in Equation 1. 

\begin{align*}
	\max_{\mathbf{x}'} \: & \mathcal{L} (\mathbf{x}', \mathbf{y}) \tag{1} \\
	s.t.  \: & \|\mathbf{x}'-\mathbf{x}\|_p \leq  \epsilon
\end{align*}
\label{eq:1}

where we denote the perturbed data, i.e., the adversarial examples, as $\mathbf{x}' \in \mathbb{R}^{n}$, adversary targeted input as $y \in \mathbb{R}$, adversarial  function as $\mathcal{L}$ which is a function of $x$, $x'$ and $y$, as well as the perturbation budget as $\epsilon$. 

The goal of the attacker is to find the optimal perturbation signal that can maximize the loss defined in Eq. 1, while at the same time keeping the perturbation magnitude small enough such that the resulting perturbed signal has invisible difference from the original signals. Adversarial function $\mathcal{L}$ and adversarial input $y$ will be tailored based on the applications. One common formulation for adversarial function is the training loss (maximize the training loss). Adversarial input is associated with adversary's goal. For classification, adversarial input can be the targeted adversarial label. For regression, adversarial input can be the targeted numerical value.

%

\textbf{Adversarial sample generation algorithm} There are several different attack generation algorithms available. In this paper we use  Basic Iterative Method (BIM) since it fits well with our attack formulation. BIM was first introduced in \cite{Kurakin16Arxiv}. It extends the FGSM method (\cite{fgsm}) into a multi-step process. The adversarial examples from the \textbf{BIM attack} can be formulated as:
\begin{equation}
{\mathbf{x}}_{0}' = {\mathbf{x}}, \quad
\mathbf{x}_{i+1}' = Clip_{\mathbf{x}, \eta} \Big\{ \mathbf{x}_{i}' + \alpha \text{sign} \big( \nabla_x J_{\theta}(\mathbf{x}_{i}', l)  \big) \Big\} \tag{2}
\label{eq.2}
\end{equation} 

where $Clip_{x, \eta} \left\{\mathbf{x}' \right\} = \min \Big\{\mathbf{x} + \eta, \max \big\{\mathbf{x} - \eta, \mathbf{x}' \big\} \Big\}$, and $\alpha$ controls the size of the update. 
Compared with FGSM, BIM attack needs multiple iterations to obtained adversarial examples. During each iteration, new $\mathbf{x}'$ will be clipped by $\eta$, which is a hyper parameter controlling the strength of the perturbation. To adapt from the image-based adversarial examples to the time series data, we remove the constrains of $\mathbf{x} \in [0,255]$ from the formulation~\cite{Kurakin16Arxiv} .


\section{Experiments and Results}
In this section we use two real PHM applications to validate the attack strategies discussed in the previous section. One of the real PHM applications is on anomaly detection and another is on fault prognostics, both of which are discussed in detail in the following two subsections, respectively.
 
\subsection{Anomaly detection}

\begin{figure}[h]
	\centering
	\includegraphics[scale=.8]{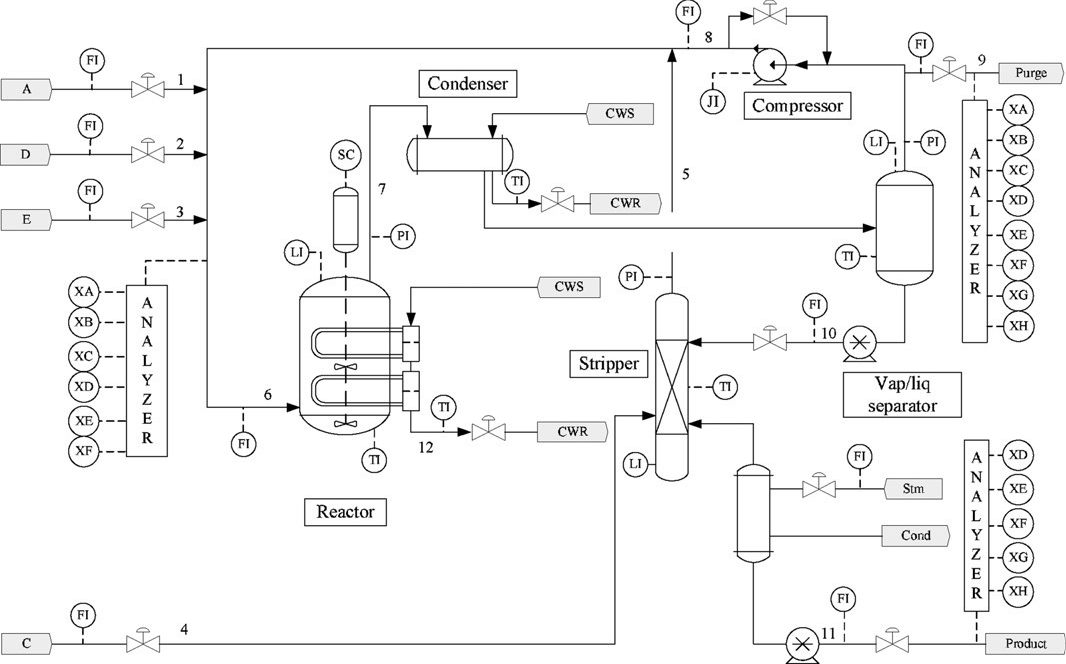}
	\caption{Tennessee Eastman Process (TEP) overall flow diagram.}
	\label{fig:TEP_flow}
\end{figure}

\textbf{The problem and the data} The Tennessee Eastman Process (TEP) is a real industrial process; and the dataset generated from the TEP simulator, a realistic simulation program of a chemical plant \cite{DOWNS1993245}, has been widely used for benchmarking fault detection algorithms/models. As the flow diagram shown in Figure \ref{fig:TEP_flow}, the process with five major units including: reactor, condenser, compressor, separator and stripper. The process has two products from four reactants. Additionally, there is an inert and a by-product. These make a total of 8 components denoted as A, B, C, D, E, F, G and H. The process has at total of 52 measurements out of which 41 are process variables and 11 are manipulated variables.
In this dataset, the system is sampled at every 3 minutes. There are 500 runs of normal operation data for training, each with 500 samples, totaling of 25 hours of operation. There are 20 process faults defined. Each of them also has 500 runs of 25 hours of operation. Testing data has similar setup except that each run has 960 samples, equivalent to 48 hours of operation. For faulty runs, fault is injected at 1-hour time step for training data, while at 8-hour time step for testing data. In this paper, we formulate the TEP anomaly detection problem as binary classification classifying normal operation from the 20 process faults.

\begin{figure}[h]
	\centering
	\includegraphics[scale=.4]{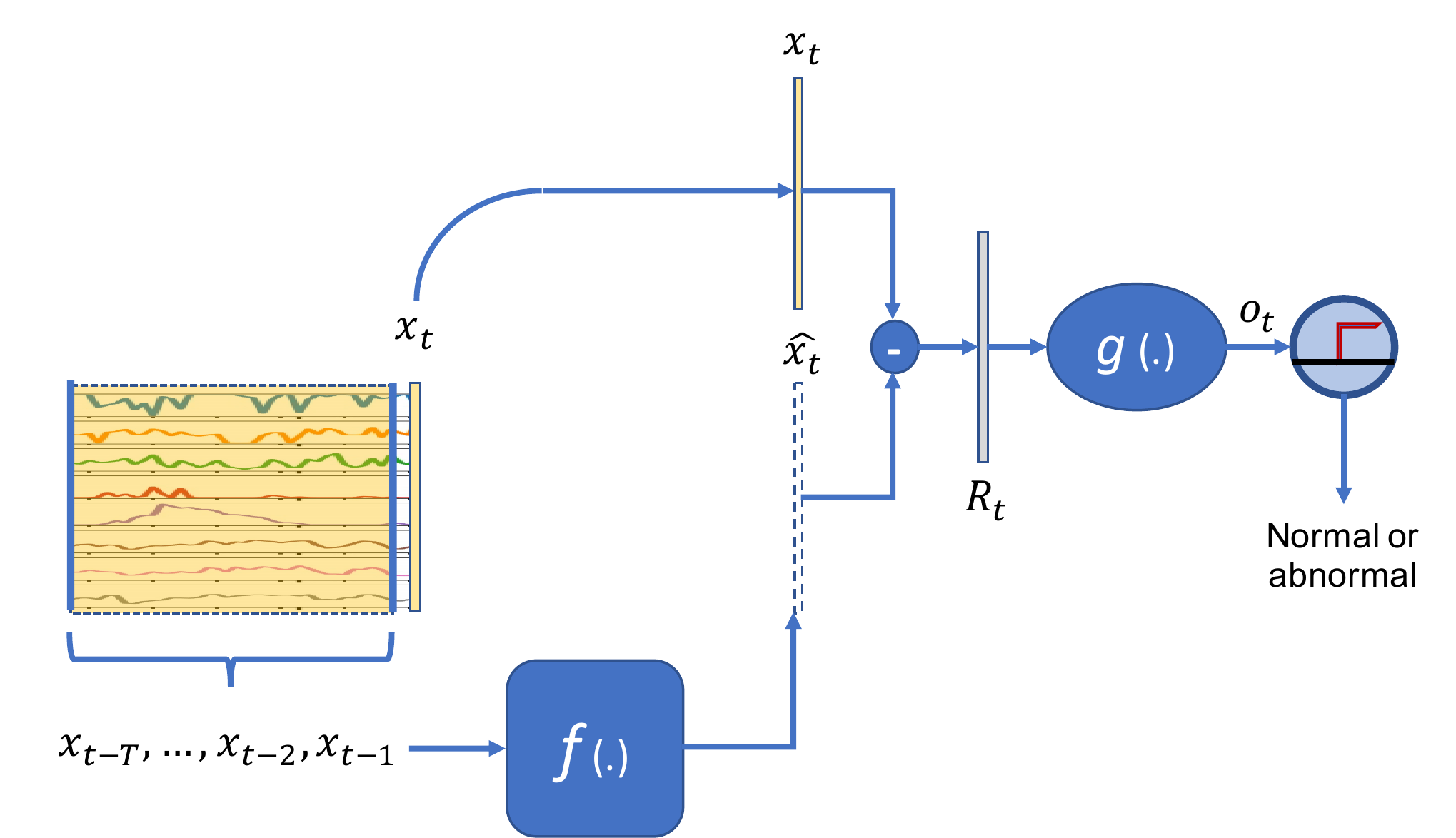}
	\caption{Flow diagram of our auto-regression residual-based anomaly detection scheme.}
	\label{fig:AD_model}
\end{figure}

\textbf{The anomaly detection model} The anomaly detection model is an auto-regression residual-based AD model. As shown in Figure \ref{fig:AD_model}, the residual-based AD scheme relies on the auto-regression model (normality model), $\mathbf{f(.)}$. In this paper, our normality model is a 2-layer stacked LSTM (Long and Short-Term Memory) with the number of hidden states of 50, which was trained using normal data only. The window length $\mathbf{T}$ (the number of lagged samples) is 120. The residual vector, $\mathbf{R_{t}}$, which is the difference between the predicted and measured values at time $t$, is then transformed to an anomaly score (a scalar), $\mathbf{o_{t}}$, by the function $\mathbf{g(.)}$, the Mahalanobis distance, to the normal data residual distribution. The anomaly score is finally thresholded to obtain the status (either normal or abnormal).

\textbf{Model performance comparison} Anomaly detection is a binary classification problem distinguishing abnormal from normal. And thus the attacker's objective for anomaly detection can fall into two categories:\\
 1) to perturb a normal sample such that the detection algorithm predicts as an abnormal sample.\\
 2) to perturb an abnormal sample such that the detection algorithm predicts as a normal sample.\\

To achieve this goal, adversarial samples to be generated for normal data need to maximize the scoring function $\mathbf{g(.)}$, while adversarial samples to be generated for abnormal data need to minimize the scoring function $\mathbf{g(.)}$. Thus, we define our adversarial function $\mathcal{L}$ as $\mathbf{g(.)}$ for normal samples and $-\mathbf{g(.)}$ for abnormal samples.

\begin{figure}[h!]
	\centering
	\includegraphics[scale=.6]{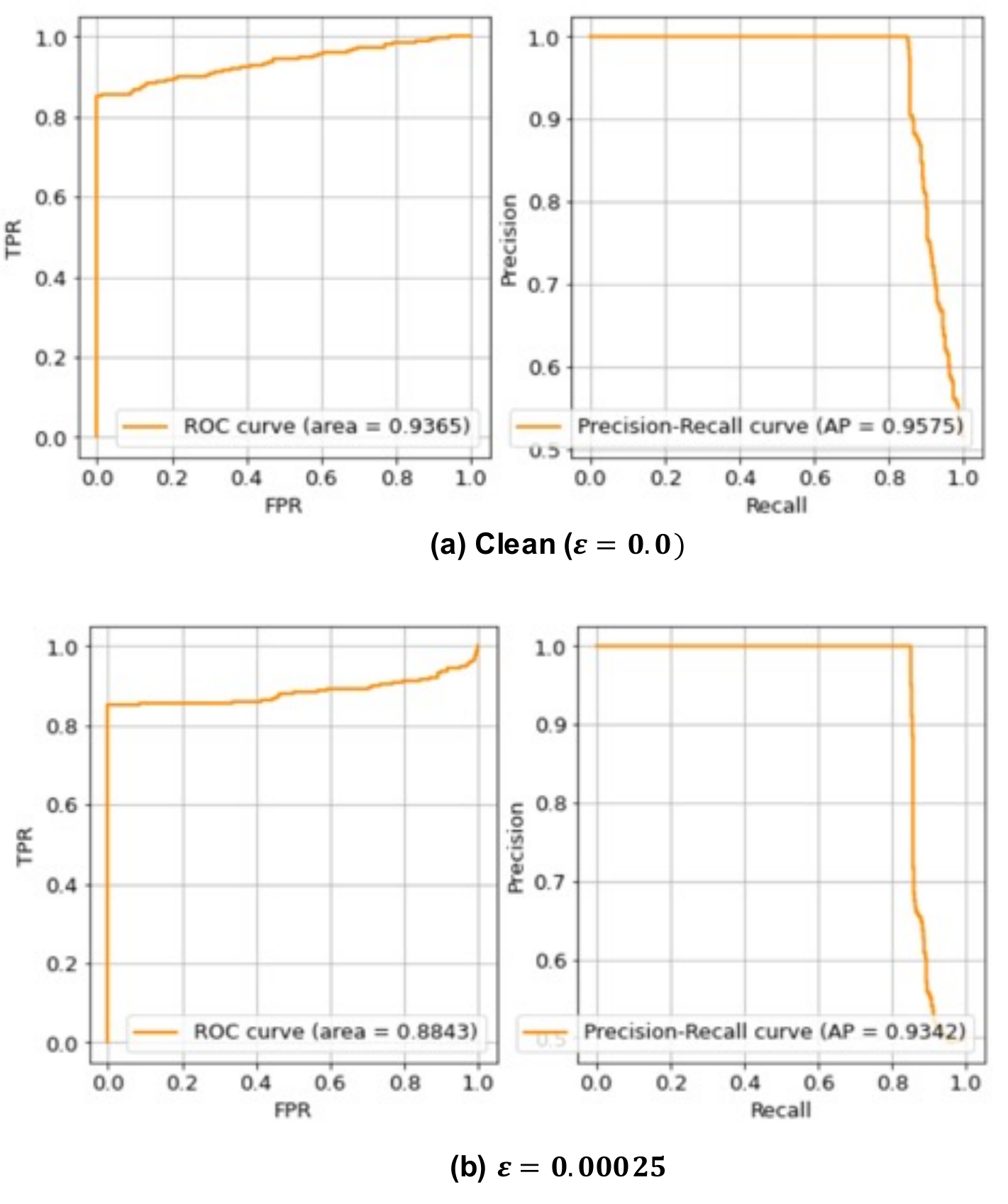}
	\caption{Comparison of ROC and PRC between (a) the clean and (b) adversarial samples with perturbation magnitude of 0.00025.}
	\label{fig:TEP_RoC}
\end{figure}

We use receiver operating characteristic (ROC) and Precision-Recall curves (PRC) as well as their associated area-under-curve ($AUC\_ROC$ and $AUC\_PRC$) as the performance metrics for anomaly detection and use the same performance metrics to demonstrate the effectiveness of adversarial examples. Figure \ref{fig:TEP_RoC} shows the comparison of the performance metrics between the clean (attack-free) model and the model subjected to the adversarial perturbation with a small perturbation magnitude of 0.00025. As can be seen from the figure, adversarial samples generated significantly degrade the performance of the detection algorithm. Specifically, the $AUC\_ROC$ is reduced from 0.9365 to 0.8843 and the $AUC\_PRC$ from 0.9575 to 0.9342.  Table \ref{table:auc_tep} also summarizes the performance metrics ($AUC\_ROC$ and $AUC\_PRC$) change as perturbation magnitude increases, which indicates that as the perturbation magnitude $\epsilon$ increases, the performance degrades significantly more.

\begin{table}[t]
	\begin{center}
	\caption{AUC-ROC and AUC-PRC under different perturbation magnitudes.}
	\label{table:auc_tep}
	\begin{tabular}{lrrr}
	\hline \hline
	Perturbation \\ Magnitude &  AUC-ROC &  AUC-PRC \\
	\hline \hline
	0.0 &   0.9365 &  0.9575 \\ \hline
	0.00025 &   0.8843 &  0.9342 \\
	0.00825 &   0.6870 &  0.8269 \\
	0.03500 &   0.6206 &  0.7821 \\ \hline
	\end{tabular}
	\end{center}
\end{table}

\begin{figure*}[t]
	\centering
	\includegraphics[scale=.76]{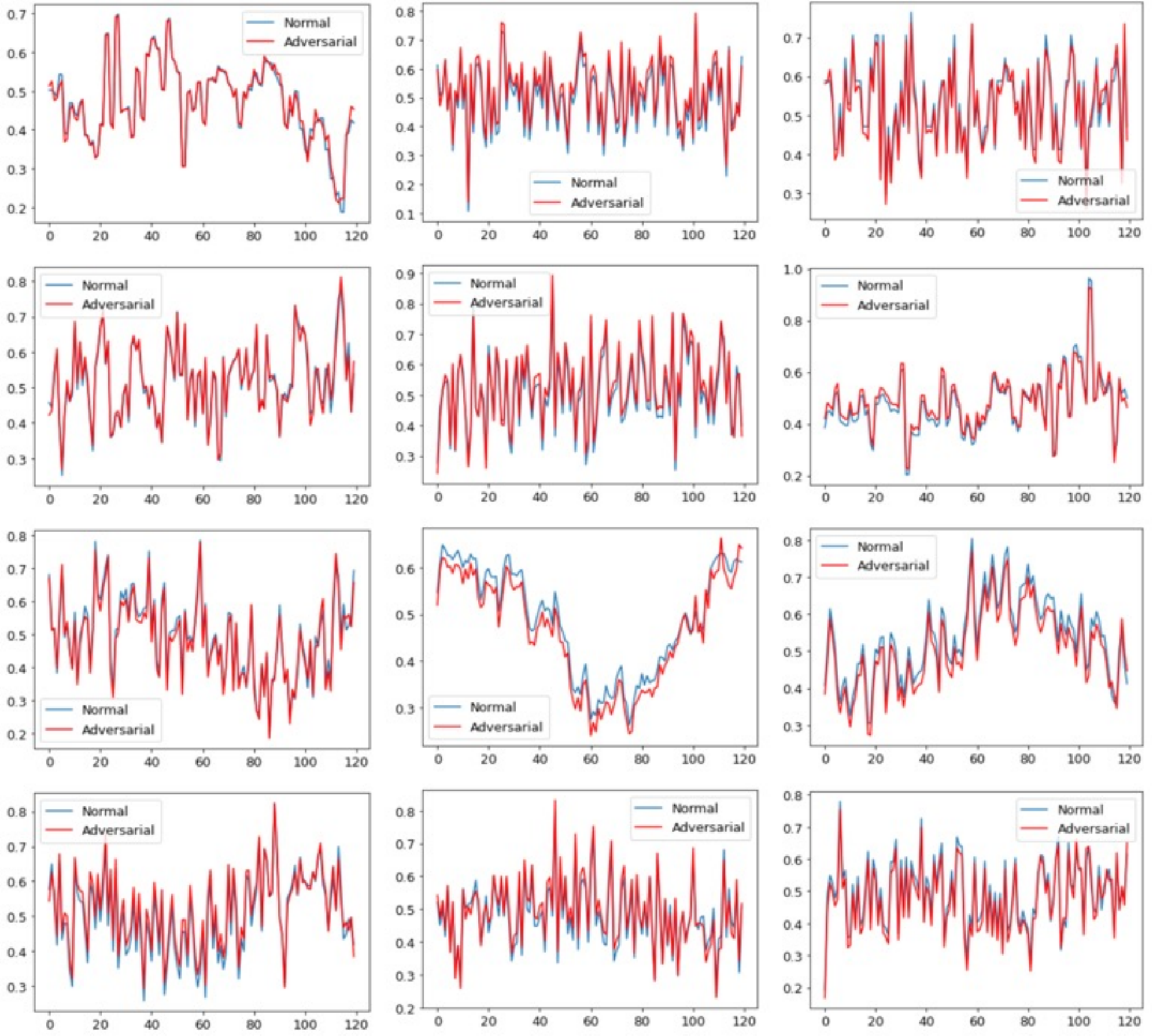}
	\caption{Comparison between clean and perturbed signals for the TEP data (perturbation magnitude of 0.035). Note: only  first 12 signals are shown. X-axis is time stamp and Y-axis is the normalized measurement.}
	\label{fig:TEP_signal}
\end{figure*}

Figure \ref{fig:TEP_signal} shows an example of comparison between a clean sample and the corresponding adversarial perturbed sample, when perturbation magnitude is 0.035. Each subplot represents a variable (sensor measurement). For plotting convenience, we randomly selected 12 variables out of the 52 variables. One can see that our perturbation is very small (almost invisible), and thus it is most likely undetectable in real world operation.

\subsection{Fault Prognosis}

\begin{figure}[h!]
	\centering
	\includegraphics[scale=.5]{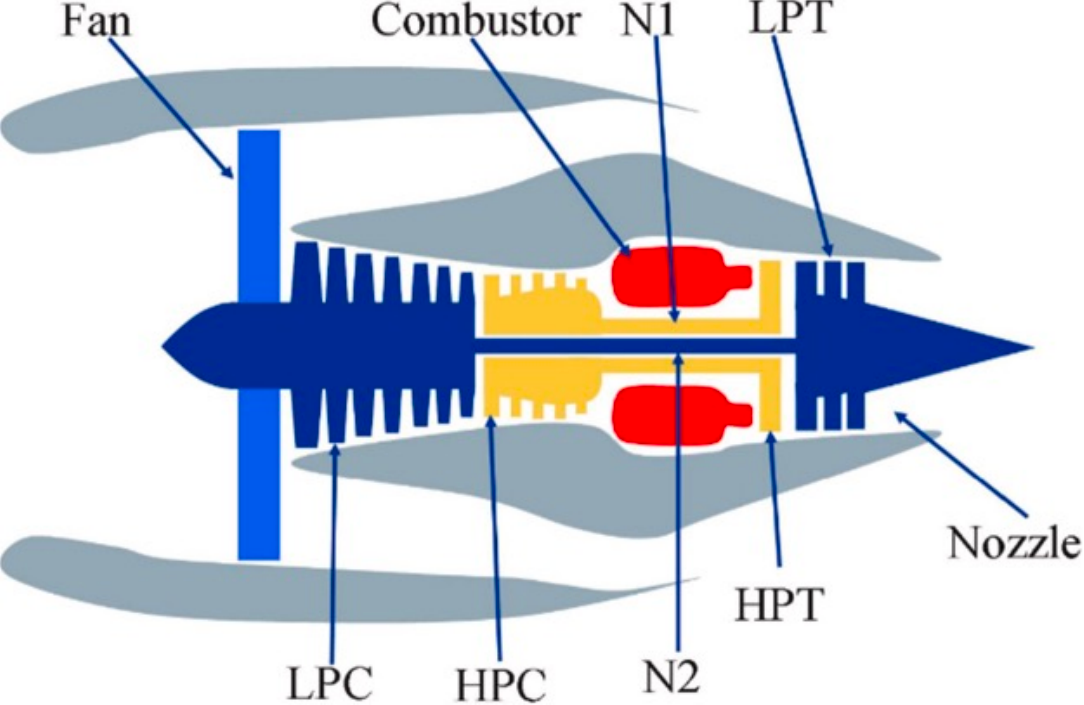}
	\caption{Turbo Fan Engine for C-MAPSS data. \\LPC = low pressure compressor; HPC=high pressure compressor; LPT=low pressure turbine; HPT = high pressure turbine; and N1, N2 = fan and core speeds.}
	\label{fig:CMAPSS}
\end{figure}

\textbf{The problem and the data}  For RUL prediction, we use the publicly available C-MAPSS datasets created by NASA \cite{Saxena2008DamagePM}, which have been popularly used in publications for benchmarking and developing prognostics algorithms. The datasets were generated using the commercial modular aero-propulsion system simulation (C-MAPSS), a turbofan engine simulation engine (see Figure \ref{fig:CMAPSS}). The C-MAPSS datasets consist of five individual datasets that differ in the number of simultaneous fault modes and the operational conditions simulated. Each of these five datasets consists of multiple multivariate time series, representing engine health status from normal to fault and to failure (i.e.,  run-to-failure data).  There are a total of 26 variables. The first 2 variables are for engine ID and timestamps. The next three variables are for defining engine operation conditions. The rest of 21 variables are the engine response measurements. In this paper we use dataset No. 3 for validating our attack strategies.

\textbf{The RUL prediction model} 
The RUL prediction (prognosis) model is to predict the remaining useful life of the engine based on the 21 measurements obtained at a given time. To perform the RUL prediction, we build a convolutional neural network (CNN) model. It consists of two convolutional layers: 19-5x17 and 25-16x1 with ReLu activation, followed by a global average pooling and a dense-connected layer. Training loss is defined as MSE between predicted RULs and targeted RULs. Time window size for the input signals is 35. We use the sliding window to augment the dataset for training of the model. The data is standardized/scaled by mean and standard deviation of individual variables. 

\begin{figure}[h]
	\includegraphics[scale=.195]{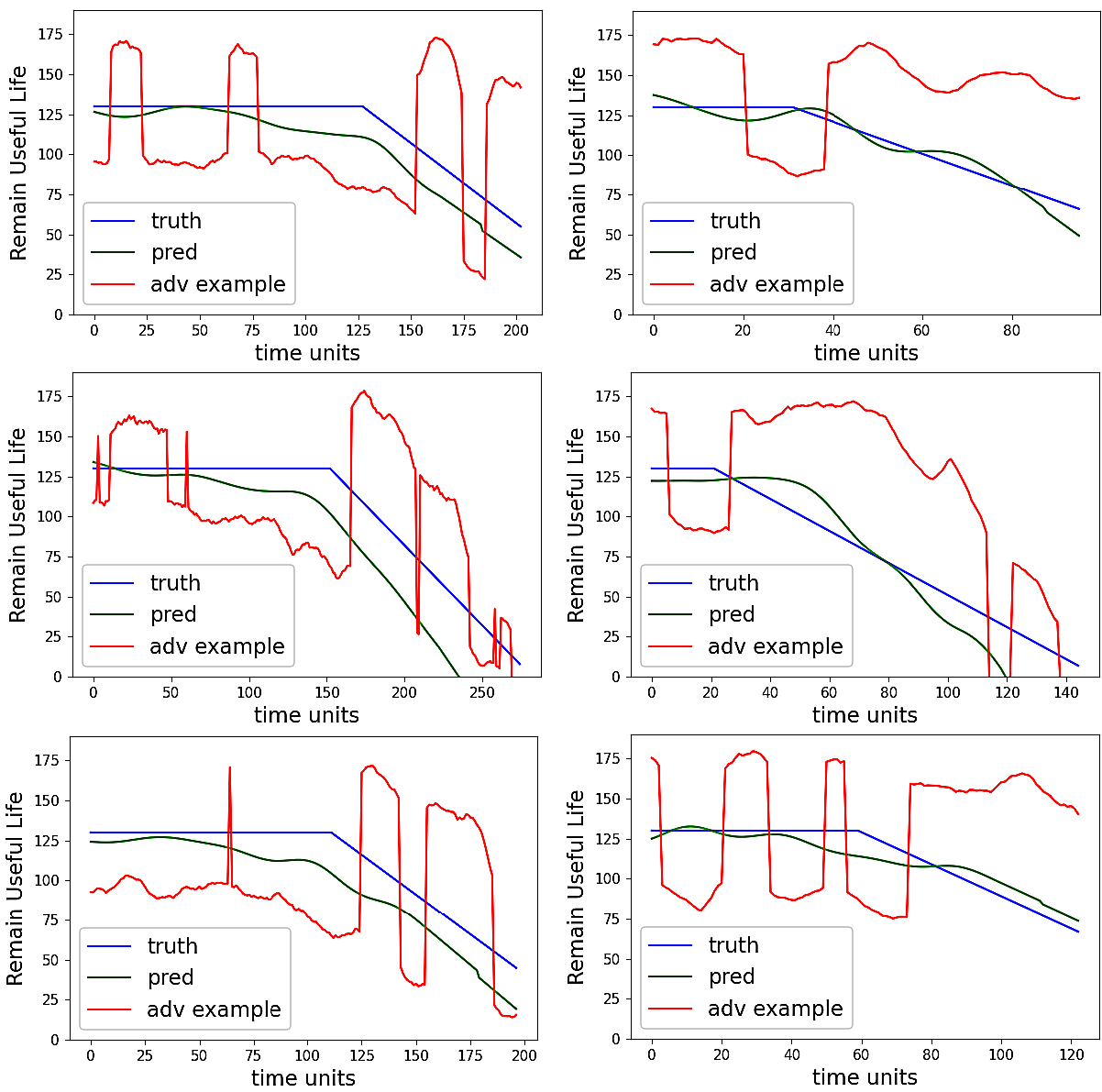}
	\caption{Comparison between the true and the predicted RULs for six samples of time-series sequences.}
	\label{fig:rul_pred_vs_true}
\end{figure}

\textbf{Model performance comparison} Prognostics or RUL prediction is a regression problem. We use MSE as the performance measure. Figure \ref{fig:rul_pred_vs_true} compares the predicted and the true RULs for six samples of time-series sequences, where for predicted RULs we show both for clean (in green color) and perturbed (in red color) model outputs. From Figure \ref{fig:rul_pred_vs_true} one can clearly see that the clean RUL prediction model performs reasonably well by tracking the RUL trajectories. And, more importantly,the adversarial sampling significantly degrades the model's prediction capability by resulting in significant increase on prediction errors.

To quantitatively show the prediction error increase due to the adversarial samples, the performance metrics(MSE) of the RUL prediction models between the clean model and the model under adversarial sampling with different perturbation magnitudes are shown in Table \ref{table:mse_vs_epislon}. With a small perturbation magnitude of 0.025, the MSE increases from 242.93 of the clean model to 421.05 after the perturbation, a 73.3\% increase on MSE. Such prediction error increase can result in a significant economic consequences. For example, over-prediction of RUL could lead to missing a timely maintenance which might cause a catastrophic damage to the asset monitored. 

\begin{table}[t]
	\begin{center}
	\caption{Performance (MSE) of the prediction models with different perturbation magnitudes.}
	\label{table:mse_vs_epislon}
	\begin{tabular}{lrr}
	\hline \hline
	Perturbation \\ Magnitude &  MSE \\
	\hline \hline
	0.000 &   242.93 \\ \hline
	0.025 &   421.05 \\
	0.045 &   876.89 \\
	0.065 &   1504.13 \\ \hline
	\end{tabular}
	\end{center}
\end{table}
 
\begin{figure*}[t]
	\centering
	\includegraphics[scale=.66]{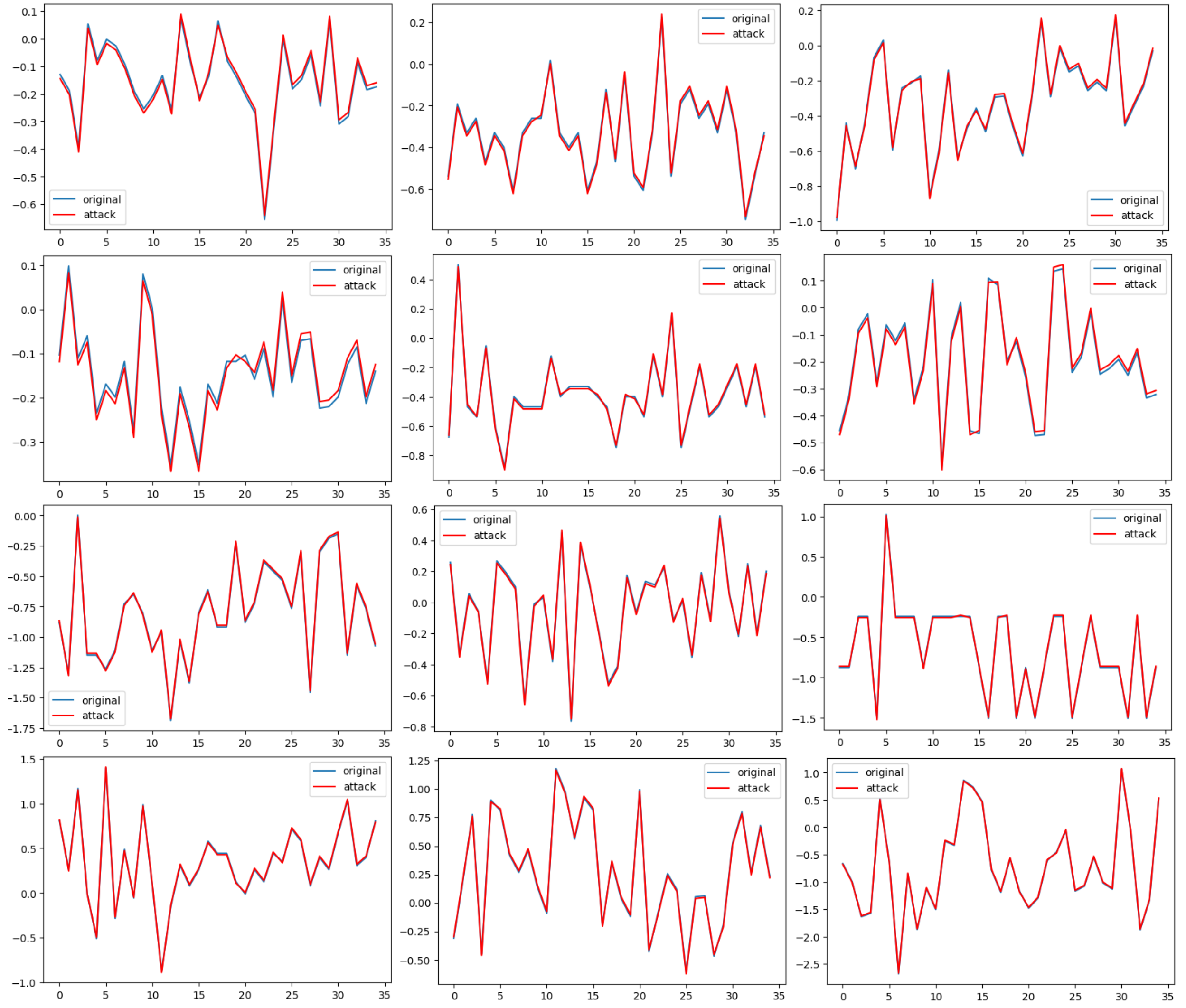}
	\caption{Comparison between clean and perturbed signals for the C-MAPSS data (with perturbation magnitude of 0.025). X-axis is the time unit and Y-axis is the normalized variable value (measurement).}
	\label{fig:CMAPSS_signal}
\end{figure*}

Figure \ref{fig:CMAPSS_signal} shows comparison between a clean sample and the corresponding adversarial perturbed sample, when perturbation magnitude is 0.025. Each subplot shows a normalized variable (sensor measurement) over time (cycle). For plotting convenience, we only show 12 variables out of the 21 variables. One can see that our perturbation is very small (almost invisible), and thus it is most likely undetectable in real world operation.

\section{Conclusion}
The vulnerability of deep learning models to adversarial attacks has been actively studied in the domains of CV and NLP. PHM algorithms' vulnerability to adversarial attacks, however,  has not yet attracted too much attention in the PHM community, despite the fact that deep learning models have been increasingly adopted to PHM applications. This paper presents an initial study on PHM algorithms' vulnerability to adversarial attacks. We discussed the strategies of attacking PHM algorithms by considering their unique characteristics of PHM data type, i.e., time-series sensor measurements. Experiments on the two real-world PHM applications case studies validated the effectiveness of the attack strategies and demonstrated that PHM algorithms indeed are vulnerable to adversarial attacks.

Our future work will include: \\
1) Investigating other two types of attacks, namely black-box and gray-box attacks, to PHM algorithms in addition to the white-box attacks considered in this paper. \\
2) Exploring strategies for attacking traditional machine learning (i.e., non DL) PHM algorithms. \\
3) Exploring strategies on defending adversarial attacks to PHM algorithms.


\section*{Nomenclature}

\begin{tabular}{ l  l }
	$f(.)$   &normality function \\ 
	$g(.)$    &transform function \\
	$x$	   		&multivariate time series \\ 
	$x'$   	&perturbed multivariate time series \\ 
	$y$	   		&target label \\ 
	$\mathcal{L}$ &adversarial loss function \\
	$PRC$ 		&precision-recall curves \\
	$ROC$		&receiver operating characteristic \\
	$RUL$		&remaining useful life \\
	$T$			&time window length \\
	$\epsilon$    &perturbation magnitude \\ 
 \end{tabular}

\bibliographystyle{apacite}
\bibliography{yangz, phm21_citations}

\end{document}